\documentclass[conference]{IEEEtran}

\usepackage{graphicx}  
\begin{document}

\title{\Large\bf REVISED UPPER LIMITS OF THE DIFFUSE TeV GAMMA RAYS FROM THE
 GALACTIC PLANES WITH THE TIBET II AND III AIR SHOWER ARRAYS}

\vspace{0.7mm}
\author{\authorblockN{\normalsize (The Tibet AS$\gamma$ Collaboration)\\
M. Amenomori,$^a$ S. Ayabe,$^b$ 
X.J. Bi,$^c$ D. Chen,$^d$ S.W. Cui,$^e$ 
Danzengluobu,$^f$ L.K. Ding,$^c$ X.H. Ding,$^f$ C.F. Feng,$^g$\\ Zhaoyang Feng,$^c$
Z.Y. Feng,$^h$ X.Y. Gao,$^i$ Q.X. Geng,$^i$ H.W. Guo,$^f$ H.H. He,$^c$
M. He,$^g$ K. Hibino,$^j$ N. Hotta,$^k$\\ Haibing Hu,$^f$ H.B. Hu,$^c$
J. Huang,$^l$ Q. Huang,$^h$ H.Y. Jia,$^h$ F. Kajino,$^m$ K. Kasahara,$^n$
Y. Katayose,$^d$ C. Kato,$^o$\\ K. Kawata,$^l$ Labaciren,$^f$ G.M. Le,$^p$ A.F. Li,$^g$ J.Y. Li,$^g$
Y.-Q. Lou,$^q$ H. Lu,$^c$ S.L. Lu,$^c$ X.R. Meng,$^f$ K. Mizutani,$^{b,r}$\\ 
J. Mu,$^i$ K. Munakata,$^o$ A. Nagai,$^s$ H. Nanjo,$^a$ M. Nishizawa,$^t$
M. Ohnishi,$^l$ I. Ohta,$^u$
 H. Onuma,$^b$ T. Ouchi,$^j$\\ S. Ozawa,$^l$ 
J.R. Ren,$^c$ T. Saito,$^v$ T. Y. Saito,$^l$ M. Sakata,$^m$ T. K. Sako,$^l$
T. Sasaki,$^j$ M. Shibata,$^d$ A. Shiomi,$^l$\\
 T. Shirai,$^j$ H. Sugimoto,$^w$ M. Takita,$^l$ Y.H. Tan,$^c$ N. Tateyama,$^j$
S. Torii,$^r$ H. Tsuchiya,$^x$ S. Udo,$^l$ B. Wang,$^i$\\ H. Wang,$^c$ X. Wang,$^l$
 Y.G. Wang,$^g$ H.R. Wu,$^c$ L. Xue,$^g$ Y. Yamamoto,$^m$ C.T. Yan,$^l$
 X.C. Yang,$^i$ S. Yasue,$^y$\\ Z.H. Ye,$^p$ G.C. Yu,$^h$ A.F. Yuan,$^f$ 
T. Yuda,$^j$
 H.M. Zhang,$^c$ J.L. Zhang,$^c$ N.J. Zhang,$^g$
X.Y. Zhang,$^g$\\ Y. Zhang,$^c$ Yi Zhang,$^c$ Zhaxisangzhu$^f$ and
X.X. Zhou$^h$
}
\vspace*{1.5mm}
\noindent
\authorblockA{\footnotesize $^a$Department of Physics, Hirosaki University, Hirosaki 036-8561, Japan~
$^b$Department of Physics, Saitama University, Saitama 338-8570,\\ Japan~
$^c$Key Laboratory of Particle Astrophysics, Institute of High Energy Physics, Chinese Academy of Sciences, Beijing 100049, China~\\
$^d$Faculty of Engineering, Yokohama National University, Yokohama 240-8501, Japan~
$^e$Department of Physics, Hebei Normal University,\\ Shijiazhuang
 050016, China~
$^f$Department of Mathematics and Physics, Tibet University, Lhasa 850000, China~ 
$^g$Department of Physics,\\ Shandong University,
 Jinan 250100, China~ 
$^h$Institute of Modern Physics, South West Jiaotong University, Chengdu 610031, China~\\
$^i$Department of Physics, Yunnan University, Kunming 650091, China~
$^j$Faculty of Engineering, Kanagawa University, Yokohama 221-\\8686, Japan~
$^k$Faculty of Education, Utsunomiya Universit, Utsunomiya 321-8505, Japan~
$^l$Institute for Cosmic Ray Research, University\\ of Tokyo, Kashiwa 277-8582, Japan~ $^m$Department of Physics, Konan University, Kobe 658-8501, Japan~
$^n$Faculty of Systems Engineer-\\ing, Shibaura Institute of Technology, Saitama 337-8570, Japan~
$^o$Department of Physics, Shinshu University, Matsumoto 390-8621,\\ Japan~
$^p$Center of Space Science and Application Research, Chinese Academy of Sciences, Beijing 100080, China~
$^q$Physics Department\\ and Tsinghua Center for Astrophysics, Tsinghua University, Beijing 100084, China~
$^r$Advanced Research Institute for Science and\\ Engineering, Waseda University, Tokyo 169-8555, Japan~
$^s$Advanced Media Network Center, Utsunomiya University, Utsunomiya 321-\\8585, Japan~
$^t$National Institute of Informatics, Tokyo 101-8430, Japan~
$^u$Tochigi Study Center, University of the Air, Utsunomiya 321-\\0943, Japan~
$^v$Tokyo Metropolitan College of Industrial Technology, Tokyo 116-8523, Japan~$^w$Shonan Institute of Technology, Fujisawa\\ 251-8511, Japan~ $^x$RIKEN, Wako 351-0198, Japan~
$^y$School of General Education, Shinshu University, Matsumoto 390-8621, Japan~\\
Email: yamamoto@hep.konan-u.ac.jp
}
\vspace*{-4mm}
}


%


\maketitle

\begin{abstract}
The flux upper limits of the diffuse gamma rays, from the inner and outer Galactic planes, are revised by factors of 4.0$\sim$3.7 for mode energies 3$\sim$10 TeV, respectively, by using the simulation results of the effective area ratios for gamma-ray induced showers and cosmic-ray induced ones in the Tibet air shower array. In our previous work, (Amenomori et al., ApJ, 580, 887, 2002) the flux upper limits were deduced only from the flux ratio of air showers generated by gamma rays versus cosmic rays.  The details of the simulation are given in the paper (Amenomori et al., Advances in Space Research, 37, 1932, 2006).  The present  result using the same data as in ApJ suggests that the spectral index of source electrons is steeper than 2.2 and 2.1 for the inner and outer Galactic planes, respectively.
\end{abstract}


%
\IEEEpeerreviewmaketitle

\section{Introduction}


Diffuse gamma rays in MeV$\sim$GeV energy region from the inner Galactic (IG) and outer Galactic (OG) planes observed by EGRET (Hunter et al. 1997) [1] show a sharp ridge both along the IG and OG planes. The EGRET flux is about 3 times higher than {\sl COS~B} data (Mayer-Hasselwander et al. 1982) [2] in several GeV, although the flux is consistent with the conventional calculation (Dermer 86) [3] in $E \leq$ 1 GeV. The EGRET excess above 1 GeV has been tried to explain by some models; a hard source electron spectrum of index $\beta$=2.0 by Pohl \& Esposito (1998) [4], hard proton spectra by Mori (1996) [5] and Webber (1999) [6], and an additional secondary electrons and positrons raising from the cosmic-ray collisions with ISM by Strong et al. (2004) [7]. 

In higher energy region theoretical calculations have been given by Porter \& Protheroe (1997) [8] and Tateyama \& Nishimura (2001) [9] for the inverse Compton (IC) gamma rays, and by Berezinsky et al. (1993) [10] for $\pi^0 \rightarrow 2\gamma$ process through cosmic-ray interaction with ISM. The most experimental data in higher energy region gave only flux upper limits except the definite flux by Milagro [11] at 3.5 TeV for IG. Not only the absolute flux but also the flux upper limit are both important for restriction of theoretical models. In this paper, using of the detection area ratio of the Tibet array between gamma rays and galactic cosmic rays in the simulation, we revised and decrease the flux upper limits in our previous paper (1002) [12], and compare with the Milagro result.  The details of the simulation is described in our recent paper (Amenomori et al. 2006) [13].
 

\section{Simulation of effective areas}

Shower size of primary gamma-ray induced showers is about three times larger than galactic cosmic-ray induced ones in average at the d
epth of 606 gm$^{-2}$ of the Tibet array for the multi-TeV energy region. Hence, the effective area of the array is larger for gamma rays than cosmic rays. 
Figure 1 shows the layout of the Tibet III air shower array. The Tibet II array is
\begin{figure}
\centering
\includegraphics[width=80mm]{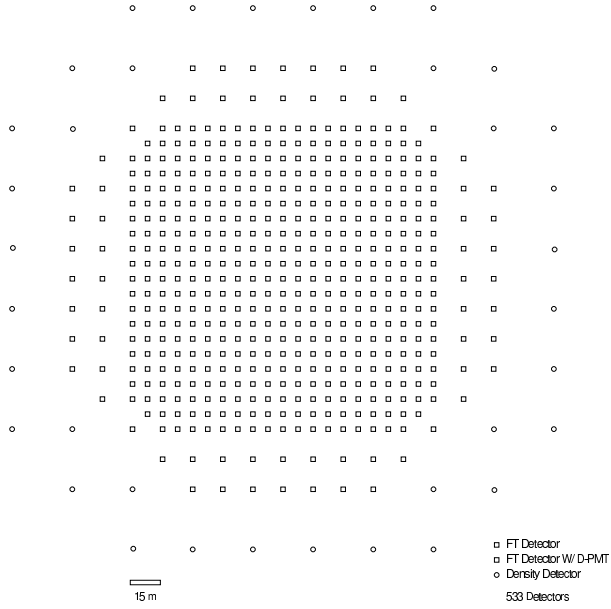}
\caption{Layout of the Tibet III array in Yangbajing at the stage in 1999 $\sim$ 2001.}
\label{fig_sim}
\end{figure}
\begin{figure}
\centering
\includegraphics[width=80mm]{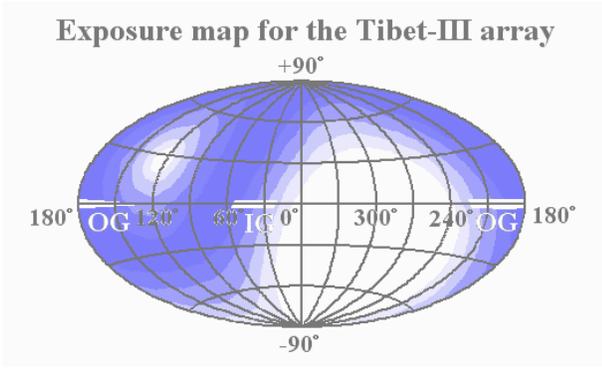}
\caption{Tibet III exposure map in the galactic
coordinates for the
 zenith angle with $\theta \leq 50^\circ$.}
\label{fig_sim}
\end{figure}

\noindent
 the one excluding detectors with the open squares.  Figure 2 is the exposure map of the Tibet III array for the zenith angle of $\leq 50^\circ$. 

Figure 3 shows differential energy spectra of triggered gamma induced showers and cosmic-ray induced ones, assuming both spectral index of 2.6 in the simulation. We can see the mode energy of triggered gamma rays is 1.5$\sim$2.0 times smaller than cosmic rays for the same trigger condition, and the effective area ratio to be about 7 in multi-TeV region.

  Figure 4 shows the average advantage factor of the effective area for gamma rays versus cosmic rays is 4.0 for $E_{\rm mode} \simeq$ 3 TeV and 3.6 for 10 TeV in average of gamma-ray spectral indices of $\beta = 2.2 \sim 2.8$. In this paper the mediate value $\beta$=2.5 is employed because of its weak dependence on the spectral index. 

The significance of the excess sigma ($\sigma$), in the Table 1, of TeV gamma rays from the IG and OG planes implies a simple

\begin{figure}
\centering
\includegraphics[width=84mm]{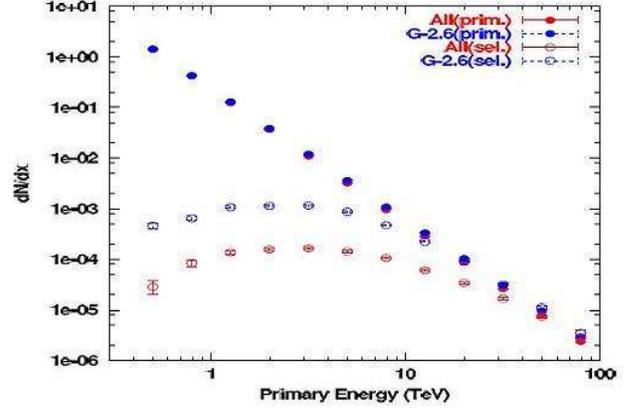}
\caption{Assumed primary spectra and distributions of triggered events
 in the simulation
 for IG plane.}
\label{fig_sim}
\end{figure}
\vspace*{6mm}
\begin{figure}
\centering
\includegraphics[width=84mm]{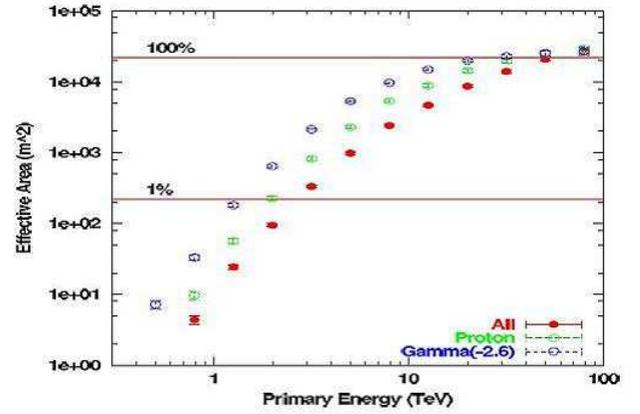}
\caption{Simulation results of effective areas of the Tibet III from IG
 plane for primary gamma rays, protons and all cosmic rays.}
\label{fig_sim}
\end{figure}
\vspace*{-6mm}

  formula of $(E-B)/\sqrt{B}$, where $E$ is the number of events on-plane
 and $B$ is the background number of events estimated from neighboring bins around the on-plane. 
Those evaluated values in the previous paper [12] are tabulated together with the simulated effective area ratio of gamma rays vs. cosmic rays and the revised upper limits in a case of small change of source electron spectral index of $\beta$ from 2.4 to 2.5.

\section{Results and Discussions}
   The present simulation, giving a larger effective area of the Tibet array for gamma rays than the galactic cosmic rays, decreases the flux upper limits of diffuse gamma rays by a significant factor, as given in Table 1. The original data in the previous paper (Amenomori et al. 2002) [12] were obtained by the Tibet III array with inner area of  22,050m$^2$ at 3 TeV, and by the Tibet II array with 28,350m$^2$ at 10 TeV.  The details of the simulation of the effective area ratio between gamma-ray/cosmic-ray initiated air showers are described for the Tibet III array, in the recent paper (Amenomori et al. 2006) [13].  In Fig. 3 is shown the detected shower event distribution in the simulation, and in Fig. 4 is also shown the effective area for


\noindent
\hspace*{9.0mm}{\small TABLE 1 Effective area ratio for gamma rays vs. cosmic rays\\
\vspace*{-1.5mm}
}\\
\vspace*{-10.5mm}
{\baselineskip=4.5mm
\hspace*{1mm}
{\footnotesize 
\begin{tabular}{|c|c|c|c|c|c|c|c|c|} \hline
Galactic latitude range   & \multicolumn{4}{c|}{$|b|\leq 2^\circ$} & \multicolumn{4}{c|}{$|b|\leq 5^\circ$}\\[0.4mm] \hline
Air shower array & \multicolumn{2}{c|}{Tibet III} & \multicolumn{2}{c|}{Tibet II} & \multicolumn{2}{c|}{Tibet III} & \multicolumn{2}{c|}{Tibet II}\\[0.3mm] \hline 
$E_{\rm mode}$ & \multicolumn{2}{c|}{3 TeV} & \multicolumn{2}{c|}{10 TeV} & \multicolumn{2}{c|}{3 TeV} & \multicolumn{2}{c|}{10 TeV}\\[0.2mm] \hline
Inner or Outer Galactic planes$^\star$ & ~~~IG~~~ & ~~~OG~~~ & ~~~IG~~~ & ~~~OG~~~ & ~~~IG~~~ & ~~~OG~~~ & ~~~IG~~~ & ~~~OG~~~ \\[0.3mm] \hline
Significance of excess ($\sigma$)$^\dagger$ & +2.52& +0.25& +1.71 & -0.63 &+1.88&+1.78&+0.81&-0.66\\[0.3mm] \hline
Flux ratio of $\gamma$ rays vs. cosmic rays$^\dagger$ &&&& &&&&\\[-0.1mm]
$I_{\gamma}(1\sigma)/I_{\rm CR}\equiv1/\sqrt{B}~(\times 10^{-4})$ & 1.95 & 1.16& 2.43 & 1.45 & 1.23 & 0.737 & 1.54& 0.936 \\ \hline
99\%CL revised flux upper limits &&&& &&&&\\
~~$J(\geq E_{\gamma})$~(10$^{-11}$cm$^{-2}$s$^{-1}$sr$^{-1}$) &57.3&19.4&7.81&2.50 &31.4&18.4&3.95&1.60\\ \hline
99\%CL upper limits ($\beta$=2.4)$^\dagger$&&&& &&&&\\
~~$E_{\gamma}^2dJ(\geq E_{\gamma})/dE_{\gamma}$ & 9.6 & 3.3 & 4.0 & 1.3 & 5.3 & 3.1 & 2.0 & 0.83\\[0.5mm]
~($\times 10^{-3}$cm$^{-2}$s$^{-1}$sr$^{-1}$MeV)&&&&&&&&\\ \hline
Effective area ratio of $\gamma$-ray/CR & \multicolumn{2}{c|}{}& \multicolumn{2}{c|}{} & \multicolumn{2}{c|}{} & \multicolumn{2}{c|}{}\\[-0.1mm]
$S_{\rm eff}(\gamma)/S_{\rm eff}$(CR) & \multicolumn{2}{c|}{4.0} & \multicolumn{2}{c|}{3.7} & \multicolumn{2}{c|}{4.0} & \multicolumn{2}{c|}{3.7} \\ \hline 
99\%CL revised upper limits ($\beta$=2.5)&&&& &&&&\\
~~$E_{\gamma}^2dJ(\geq E_{\gamma})/dE_{\gamma}$ & 2.6 & 0.88 &1.2 & 0.38 & 1.41 & 0.83 & 0.59 & 0.24\\[-0.1mm]
~($\times 10^{-3}$cm$^{-2}$s$^{-1}$sr$^{-1}$MeV)&&&&&&&&\\ \hline
\end{tabular}
}\\
\vspace*{11mm}

\noindent
{\footnotesize 
\hspace*{7mm}$\dagger$ are referred to Amenomori et al. ApJ., 580, 887 (2002)\\
\vspace*{-7.3mm}

\noindent
\hspace*{7mm}$^\star$ Ranges of IG are $20^\circ \leq l \leq 55^\circ$, and of OG $140^\circ\leq l \leq 225^\circ$}.\\
} 
\vspace*{1mm}

\noindent
 primary gamma rays, protons and all cosmic rays.

Figures 5 and 6 show the revised flux upper limits, for IG and OG planes, at $E_{\rm mode}$=3 TeV (T3: for the Tibet III array) and at 10 TeV (T2: for the Tibet II array). In these figures the EGRET data (Hunter et al. 1997) [1] of the Galactic latitude width of $|b| \leq 2^\circ$ are shown, and also shown the upper limits by Whipple (W) (LeBohec et al. 2000) [14] with 99.9\% C.L. and HEGRA (H) (Aharonian et al. 2001) [15] with 99\% C.L., though both at a small sky region around the galactic longitude of $l = 40^\circ$, and HEGRA-AIROBICC (Ha) (Aharonian et al. 2002) [16] and CASA-MIA (C-A) (Borione et al. 1998) [17] both with 90\% C.L.. Theoretical curves of inverse Compton are given by Porter and Protheroe (1997)[8] (PP2.0 and PP2.4 in figures), and by Tateyama and Nishimura (2003)[9] (TN 2.0 and TN2.4), where 2.0 and 2.4 are assumed source electron spectral indices. Theoretical curves arising from $\pi^0 \rightarrow 2\gamma$ decay are also given by Berezinsky et al. (1993)[10] (BGHS).

When the observed gamma-ray spectra with the spectral index of 2.5 is adopted in this paper, the revised results can give a strong suggestion that the spectral indices of source electrons for the inverse Compton (IC) are steeper than 2.2 in the IG plane and also 2.1 in the OG plane in comparison with the theoretical calculations of IC.

Recently, Milagro (M) obtained the definite flux at 3.5 TeV from IG plane and the upper limit from OG plane. Both are a few times lower than ours. In their IG
data are involved the range of $40^\circ \leq l \leq 100^\circ$ and $|b| \leq 5^\circ$, though our longitudinal IG range of $20^\circ \leq l \leq 55^\circ$.  Their result suggests that the diffuse gamma rays from the Galactic plane can be interpreted not only by IC but also by $\pi^0 \rightarrow 2\gamma$ below 10 TeV. The longitudinal range of Milagro is larger than our range of 20$^\circ \leq l \leq 55^\circ$ and the number of events is 3.6 times larger than ours for IG with $|b| \leq 5^\circ$. The advantage factor of hadron rejection

\hspace*{-5.5mm}{\small and gamma-ray flux upper limits.}
\vspace*{63mm}

\begin{figure}[h]
\centering
\includegraphics[width=69mm]{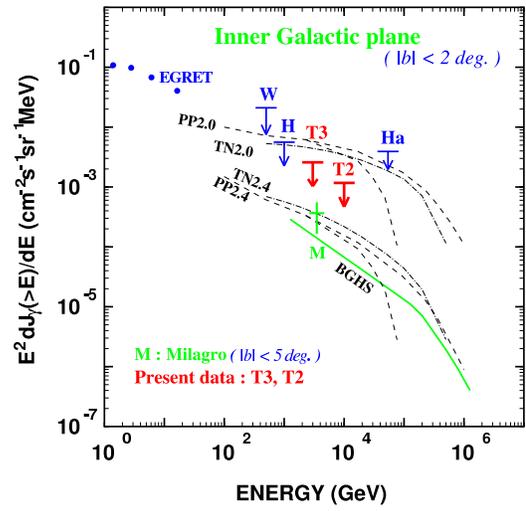}
\caption{Diffuse gamma rays from the IG plane of 20$^\circ \leq l \leq 55^\circ$ for the belt with $|b| \leq 2^\circ$.}
\label{fig_sim}
\end{figure}
\begin{figure}[h]
\centering
\includegraphics[width=69mm]{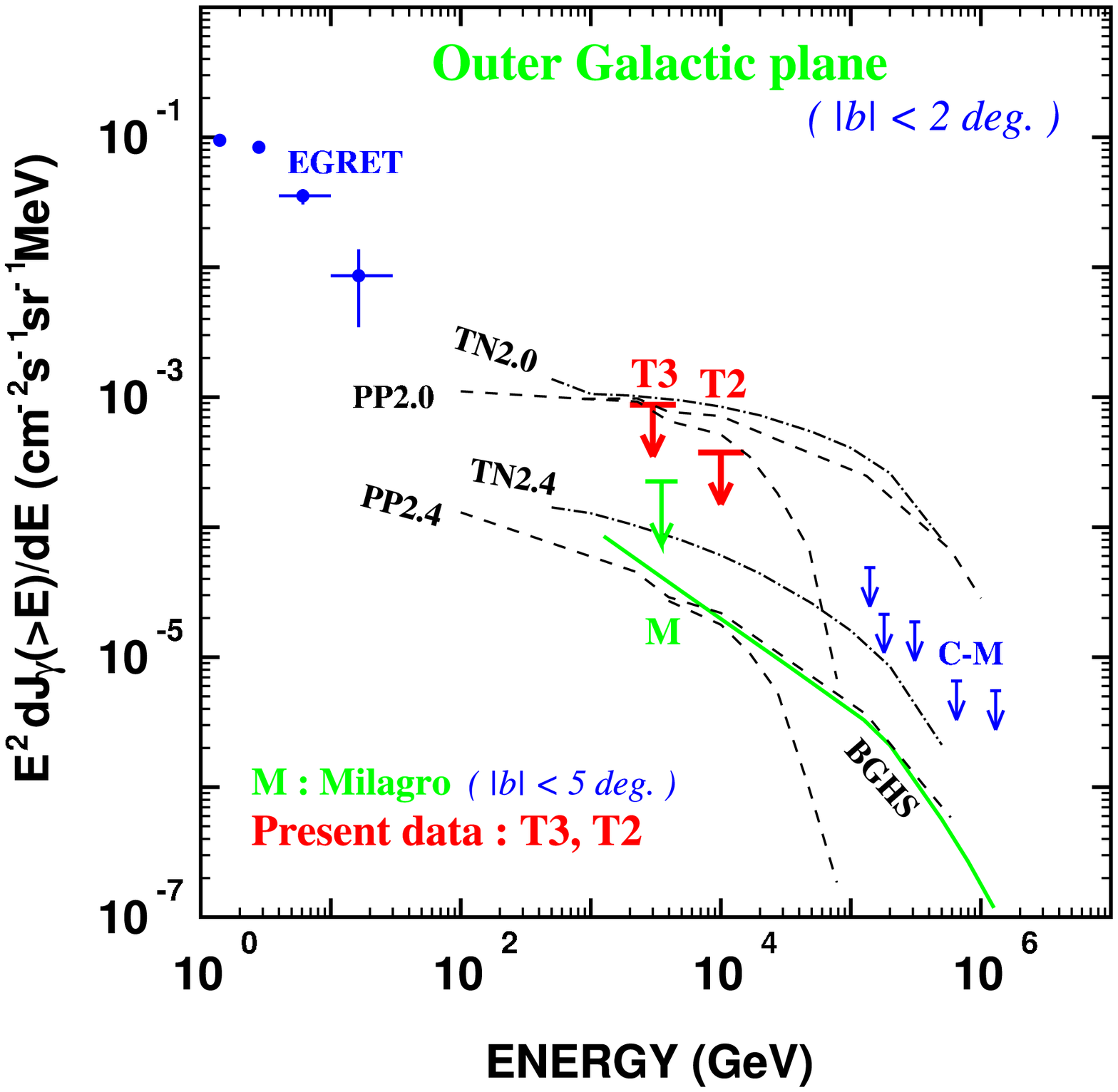}
\caption{Diffuse gamma rays from the OG plane
 of 140$^\circ \leq l \leq 225^\circ$
 for the belt with $|b| \leq 2^\circ$.}
\label{fig_sim}
\end{figure}

\begin{figure}
\centering
\hspace*{36mm}
\includegraphics[width=91.6mm]{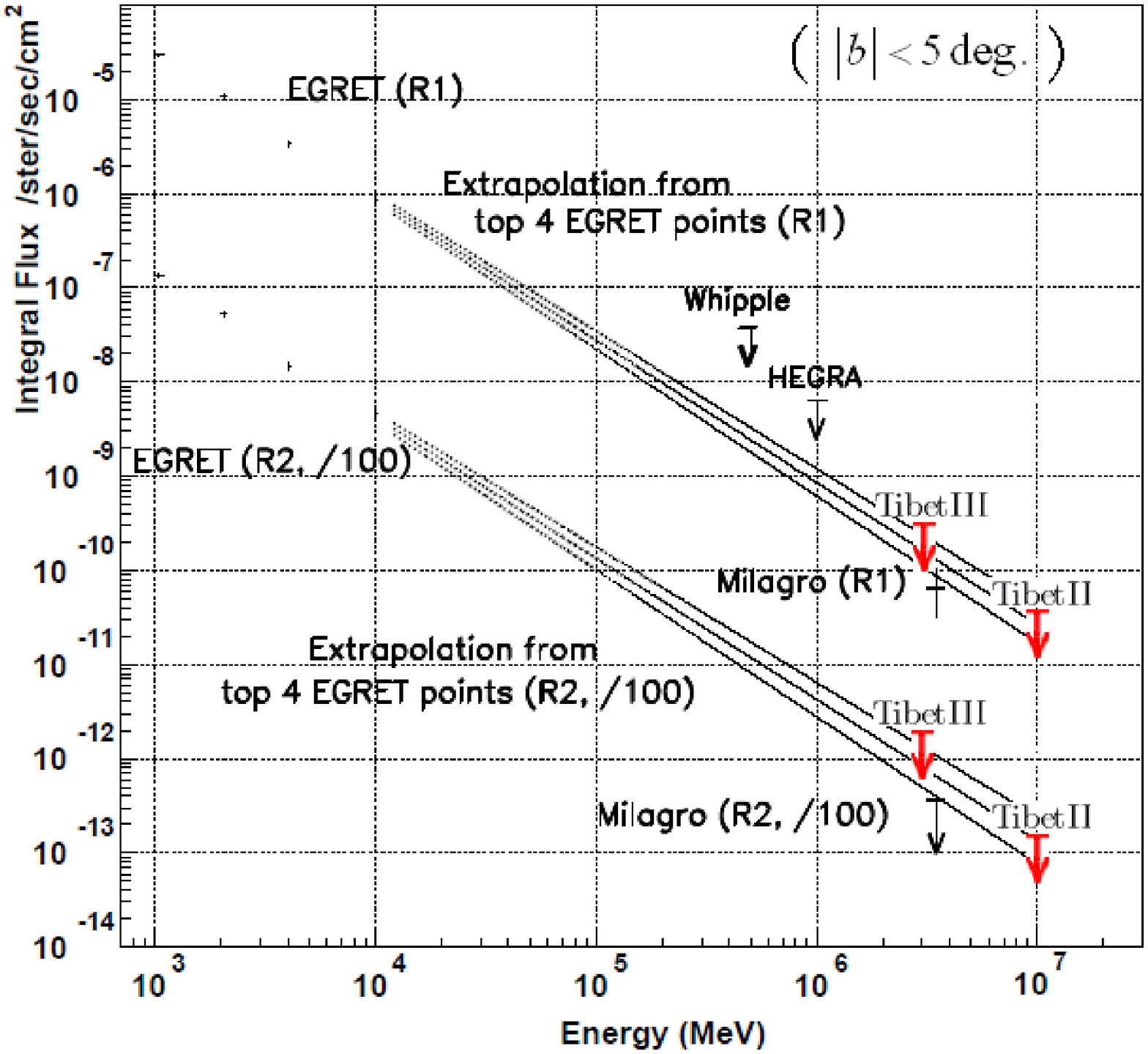}
\label{fig_sim}
\end{figure}
\vspace*{-16mm}
\hspace*{40mm}{\footnotesize Fig. 7~~Integral flux upper limits of
\hspace*{44mm}on the figure referred from}

\vspace*{8mm}
\noindent
 of Milagro is estimated to be $0.45/\sqrt{0.1}\sim 1.45$. Thus the effective area ratio of detection of gamma-ray showers vs. hadron showers is 1.45$\sqrt{3.6} \sim 3$ times better than our 99\%C.L. upper limit corresponding to 4 $\sim 5 \sigma$. Their prots of Milagro data are lower by a factor of about 4 than ours. This might be due to an odd analysis method of "time swapping". Otherwise the effective area estimated by Milagro is large. 

It seems that the recent Milagro result [11] rather claims $\pi^0 \rightarrow 2\gamma$ process by cosmic-ray particles according to the interpretation that their data is on the straight line extrapolated from the data above 10 GeV of EGRET (Hunter et al. 1997)[1].  Figure 7 is referred from Milagro paper (Atkins et al. 2005) [11] and includes our present revised upper limits for $|b| \leq 5^\circ$. This range of the Galactic plane covers the Cygnus region in which a lot of complicated point sources or diffuse gamma-ray sources seem to be involved. 
This is the reason why we have excluded this region from the analysis of the inner Galactic plane, and we have been examining the Cygnus region separately from IG plane. 

Recently, we reported our anisotropy result for the multi-TeV galactic 
cosmic rays in the northern sky [18]. In that paper, a broad hill excess of
cosmic-ray intensity is found in the Cygnus region in 13.3$\sigma$, and clear 
excesses are observable for different energy ranges in the same region. 
 
\section{Conclusions}

The Tibet AS$\gamma$ data collected between 1997 and 2001 have been reanalyzed with improved background efficiency calculation by about a factor of 4. Assuming an index 
of 2.5 for gamma-ray spectrum when calculating gamma-ray efficiency, 99\%C.L. flux limits are derived which suggests that the corresponding source electron spectrum for the inverse Compton (IC) is steeper than 2.2 for the IG plane and 2.1 

\hspace*{136mm}
\begin{minipage}[t]{10mm}
\vspace*{80.0mm}
\end{minipage}
\hspace*{-2.5mm}{\footnotesize Tibet data with $|b| \leq 5^\circ$ plotted\\
\hspace*{-16mm}the Milagro's paper PRL, 95, 251103(2005).}\\
\vspace*{4mm}

\noindent
for the OG plane. However, those gamma-ray flux limits are 
too high to constrain proton source models. In the near future, the broad 
hill of excess in the Cygnus region will be further studied with more available 
statistic. As the current Tibet air shower array is not able to discriminate 
the gamma-ray induced air shower from the hadron-induced one, we are 
planning to add dozens of large water cherenkov detectors in the Tibet array, 
to reject the hadron initiated air showers by using the information of secondary muon. Our gamma ray sensitivity is 
foreseen to be improved by a factor of 4$\sim$10 at 10$\sim$100 TeV, while at 
PeV energy we will be free from background [19].


\vspace*{4mm}
\begin{thebibliography}{1}

\bibitem{IEEEhowto:kopka}
H.~Mayer-Hasselwander et al., \emph{A\&A, 105, 164}, 1982.
\bibitem{IEEEhowto:kopka}
S. D.~Hunter et al., \emph{ApJ, 481, 205}, 1997.
\bibitem{IEEEhowto:kopka}
C. D. Dermer, \emph{A\&A, 157, 223}, 1986.
\bibitem{IEEEhowto:kopka}
M. Pohl and J. A. Esposito, \emph{ApJ, 507, 327}, 1998.
\bibitem{IEEEhowto:kopka}
M. Mori, \emph{ApJ, 478, 225}, 1997.
\bibitem{IEEEhowto:kopka}
W. R. Webber, \emph{Proc. 26th Int. Cosmic-Ray Conf. (Salt Lake City), 4, 97}, 1999.
\bibitem{IEEEhowto:kopka}
A. W. Strong, I. V. Moskalenko and O. Reimer, \emph{ApJ, 613, 962}, 2004.
\bibitem{IEEEhowto:kopka}
T. A. Porter and R. J. Protheroe, \emph{J. Phys. G, 23, 1765}, 1997.
\bibitem{IEEEhowto:kopka}
N. Tateyama and J. Nishimura, \emph{Proc. 28th Int. Cosmic-Ray Conf. (Tsukuba), 4, 2285}, 2003.
\bibitem{IEEEhowto:kopka}
V. S. Berezinsky et al., \emph{Astropart. Phys., 1, 281}, 1993.
\bibitem{IEEEhowto:kopka}
R. Atkins et al., \emph{PRL, 95, 251103}, 2005.
\bibitem{IEEEhowto:kopka}
M. Amenomori et al., \emph{ApJ, 580, 887}, 2002.
\bibitem{IEEEhowto:kopka}
M. Amenomori et al., \emph{Advances in Space Research, 37, 1932}, 2006.
\bibitem{IEEEhowto:kopka}
S. LeBohec et al., \emph{ApJ, 539, 209}, 2000.
\bibitem{IEEEhowto:kopka}
F. A. Aharonian et al., \emph{A\&A, 375, 1008}, 2001.
\bibitem{IEEEhowto:kopka}
F. A. Aharonian et al., \emph{Astropart. Phys., 17, 459}, 2002.
\bibitem{IEEEhowto:kopka}
A. Borione et al., \emph{ApJ, 493, 175}, 1998.
\bibitem{IEEEhowto:kopka}
M. Amenomori et al., \emph{SCIENCE 314, 20 October, 439}, 2006.
\bibitem{IEEEhowto:kopka}
M. Amenomori et al., Astrophysics and Space Science; Proc. Multi-Messenger Approach to High Energy Gamma Ray Sources, Barcelona,  2006 \ in press; see astro-ph/0611030.

\end{thebibliography}
\end{document}